# Observation of Emergent Superconductivity in the Quantum Spin Hall Insulator Ta$_2$Pd$_3$Te$_5$ via Pressure Manipulation

*Hui Yu*[1,2,†], *Dayu Yan*[1,†], *Zhaopeng Guo*[1], *Yizhou Zhou*[1,2], *Xue Yang*[1,2], *Peiling Li*[1], *Zhijun Wang*[1], *Xiaojun Xiang*[1,2], *Junkai Li*[4], *Xiaoli Ma*[1], *Rui Zhou*[1,2,3], *Fang Hong*[1], *Yunxiao Wuli*[1,2], *Youguo Shi*[1,3]\*, *Jian-Tao Wang*[1,2,3]\*, *Xiaohui Yu*[1,2,3]\*

[1]Beijing National Laboratory for Condensed Matter Physics, Institute of Physics, Chinese Academy of Sciences, Beijing 100190, China

[2]School of Physical Sciences, University of Chinese Academy of Sciences, Beijing 100049, China

[3]Songshan Lake Materials Laboratory, Dongguan, Guangdong 523808, China

[4]Center for High Pressure Science and Technology Advanced Research, Beijing, 100094, P.R. China

\*Correspondence and requests for materials should be sent to Y. G. S. (email: ygshi@aphy.iphy.ac.cn), J. T. W. (email: wjt@aphy.iphy.ac.cn) and X. H. Y. (email: yuxh@iphy.ac.cn)

†These authors contributed equally to this work.

Keywords: topological insulator, synchrotron X-ray diffraction, superconductivity, high pressure.

**Abstract.** Quantum Spin Hall (QSH) insulators possess distinct helical in-gap states, enabling their edge states to act as one-dimensional conducting channels when backscattering is prohibited by time-reversal symmetry. However, it remains challenging to achieve high-performance combinations of nontrivial topological QSH states with superconductivity for applications and requires understanding of the complicated underlying mechanisms. Here, our experimental observations for a novel superconducting phase in the pressurized QSH insulator Ta$_2$Pd$_3$Te$_5$ is reported, and the high-pressure phase maintains its original ambient pressure lattice symmetry up to 45 GPa. Our *in-situ* high-pressure synchrotron X-ray diffraction, electrical transport, infrared reflectance, and Raman spectroscopy measurements, in combination with rigorous theoretical calculations, provide compelling evidence for the association between the superconducting behavior and the abnormal densified phase. The isostructural transition was found to modify the topology of the Fermi surface directly, accompanied by a fivefold amplification of the density of states at 20 GPa compared to ambient pressure, which synergistically fosters the emergence of robust superconductivity. A profound comprehension of the fascinating properties exhibited by the compressed Ta$_2$Pd$_3$Te$_5$ phase is achieved, highlighting the extraordinary potential of van der Waals (vdW) QSH insulators for exploring and investigating high-performance electronic advanced devices under extreme conditions.

# 1. Introduction

Symmetry-protected topological insulators (TI) hold immense promise for revolutionizing various fields, including quantum information processing, nanoelectronics, and energy technologies, by harnessing the exceptional properties of topologically protected states and enabling unprecedented advancements in these domains. Quantum Spin Hall (QSH) topological insulators are materials with insulating bulk but conductive states on their edges due to the specific topology of their electronic band structures.[1–3] The topological protection in two-dimensional (2D) TIs arises from the nontrivial topology of their band structures.[4–7] Quantum states associated with these topological features are robust against certain types of disturbances and decoherence,[8–10] rendering the implementation of topological qubits based on 2D TIs more resilient to errors, and thus potentially more reliable for practical quantum computing.[6,11–16] Furthermore, the conductive states on the edges or surfaces of 2D TIs can be used to separate charge carriers and reduce recombination losses in photovoltaic devices. Another potential application of 2D TIs is in spintronics, where they could be used to generate and control spin currents. Weak and non-covalent van der Waals (vdW) forces dominate the interlayer interactions in vdW materials, the unique characteristic that enables the prefabrication of high-quality building blocks with desired optical properties. The assembly of these building blocks is achieved by vdW interactions, which do not impose any constraints on lattice parameters, crystal structures, or orientations.[17,18] Consequently, by combining different layers of vdW materials with complementary optical properties, it is possible to design, and fabricate integrated photonic devices with tailored functionalities, such as waveguides, modulators, detectors, and light emitters,[19–21] enabling researchers to study polaron physics in controlled environments. For example, the strong planar anisotropy, optical dichroism, and tunable absorption response of $Ta_2Pd_3Se_8$ can be integrated for ultracompact chip-integrated spectrometers, optical polarizers, and modulators applications. As research in this realm continues to make strides, it becomes urgent to explore novel materials and devise innovative techniques for the effective realization and manipulation of QSH states.

Recent theoretical works have proposed that the bulk $Ta_2Pd_3Te_5$ behaves as topological insulator characterized by relatively subtle interlayer vdW interactions.[22] Angle-resolved photoemission spectroscopy revealed the existence of an inverted band gap at the Fermi level in bulk $Ta_2Pd_3Te_5$. Moreover, the previous scanning tunneling microscopy study demonstrated the electronic structure of $Ta_2Pd_3Te_5$ has a 43-meV gap,[23] while its topological edge states were found to be similar to other topological materials such as $TaIrTe_4$, $Na_3Bi$, and $ZrTe_5$.[24–27] The investigation of atomic monolayers in vdW QSH insulator materials has ignited substantial interest due to their tunable properties and versatile functionalities, such as anomalous Hall effect,[28–30] quantum valley Hall,[31,32] excitonic insulator, and Majorana zero modes.[7,11,16] High-pressure measurements, serving as an efficient and pristine manner to finely tune the properties of both the lattice and electronic states, specifically within the intricate realm of quantum states.[33–38] Indeed, pressure-induced superconductivity and topological phase transitions have been experimentally observed in the analogous compounds of Ta$_2$NiSe$_5$[39,40] and Ta$_2$Ni$_3$Te$_5$.[41] Distinguishing itself from Ta$_2$Ni$_3$Te$_5$, its remarkable nontrivial mirror Chern number in the $k_y$ = 0 plane, due to band inversion at Γ point, ensures that Ta$_2$Pd$_3$Te$_5$ is not a trivial insulator but rather a rotation-protected topological insulator at ambient pressure, rendering it desirable for further experimental studies under high pressure.[42]

In this work, we report the comprehensive study of two isostructural phase transitions and emerging superconductivity in pressurized Ta$_2$Pd$_3$Te$_5$ by performing detailed measurements of electrical resistance, optical infrared reflectance, *in-situ* high-pressure synchrotron X-ray diffraction, and Raman spectroscopy. First-principles calculations reveal that the compact phase of Ta$_2$Pd$_3$Te$_5$ characterized by *Pnma* symmetry at 45GPa, and its volume show 16.12% reduction than that of the ambient pressure phase, manifesting its elevated state density at the Fermi level. Notably, kinetic phase conversion calculations demonstrate that the establishment of Te-Te bonds across bilayers poses a formidable obstacle to the generation of balanced high-pressure frameworks, and the pristine phase (*Pnma*) exhibits superior energetic stability compared to the possible monoclinic phase (*P*2$_1$/*m*) and layered sliding orthorhombic phase (*Pnnm*).

## 2. Results and discussion

### 2.1. Transfer from 2D vdW quantum spin hall insulator to superconductor

According to the schematic diagram presented in **Figure 1**a, Ta$_2$Pd$_3$Te$_5$ exhibits a layered van der Waals structure and crystallizes in orthorhombic arrangement with the space group *Pnma* (No.62), as confirmed by the powder X-ray diffraction (XRD) results shown in Supporting Information Figure S1. Single-crystal XRD measurements illustrate that Ta$_2$Pd$_3$Te$_5$ features preferential lattice orientations [L00], as depicted in Figure 1c. The lattice parameters determined from XRD data are $a$ = 13.9531(6) Å, $b$ = 3.7038(2) Å, and $c$ = 18.5991(8) Å, which are consistent with the findings of previous studies.[23] The SEM image (Figure 1d) indicates that majority of sample chunks are single crystals. Furthermore, the chemical composition determined by EDS approximated to the expected chemical stoichiometry. The EDS results (Table S1) demonstrate that all samples have a uniform distribution, suggesting the absence of additional phases. The temperature-dependent behavior of electrical resistivity at ambient conditions is presented in Figure 1e. The $R(T)$ data could be effectively fitted using the Arrhenius model $\rho \sim \exp(E_{act}/k_BT)$, indicating that bulk Ta$_2$Pd$_3$Te$_5$ manifests narrow-gap semiconductor features with an estimated activation energy of 30 meV. The monolayer Ta$_2$Pd$_3$Te$_5$ is constructed from stacked chains of pyramid-like TaTe$_5$ units, wherein each Ta atom is coordinated to five Te neighbors. These chains of TaTe$_5$ are interspersed with distinct interstitial sites that can accommodate one A site and two B sites filled with Pd atoms. The atomic sites are listed in Table S2. Notably, the Pd coordination to chalcogenides in this compound is tetrahedral, which contrasts with the square-planar configuration commonly observed in most four-coordinate Pd complexes.

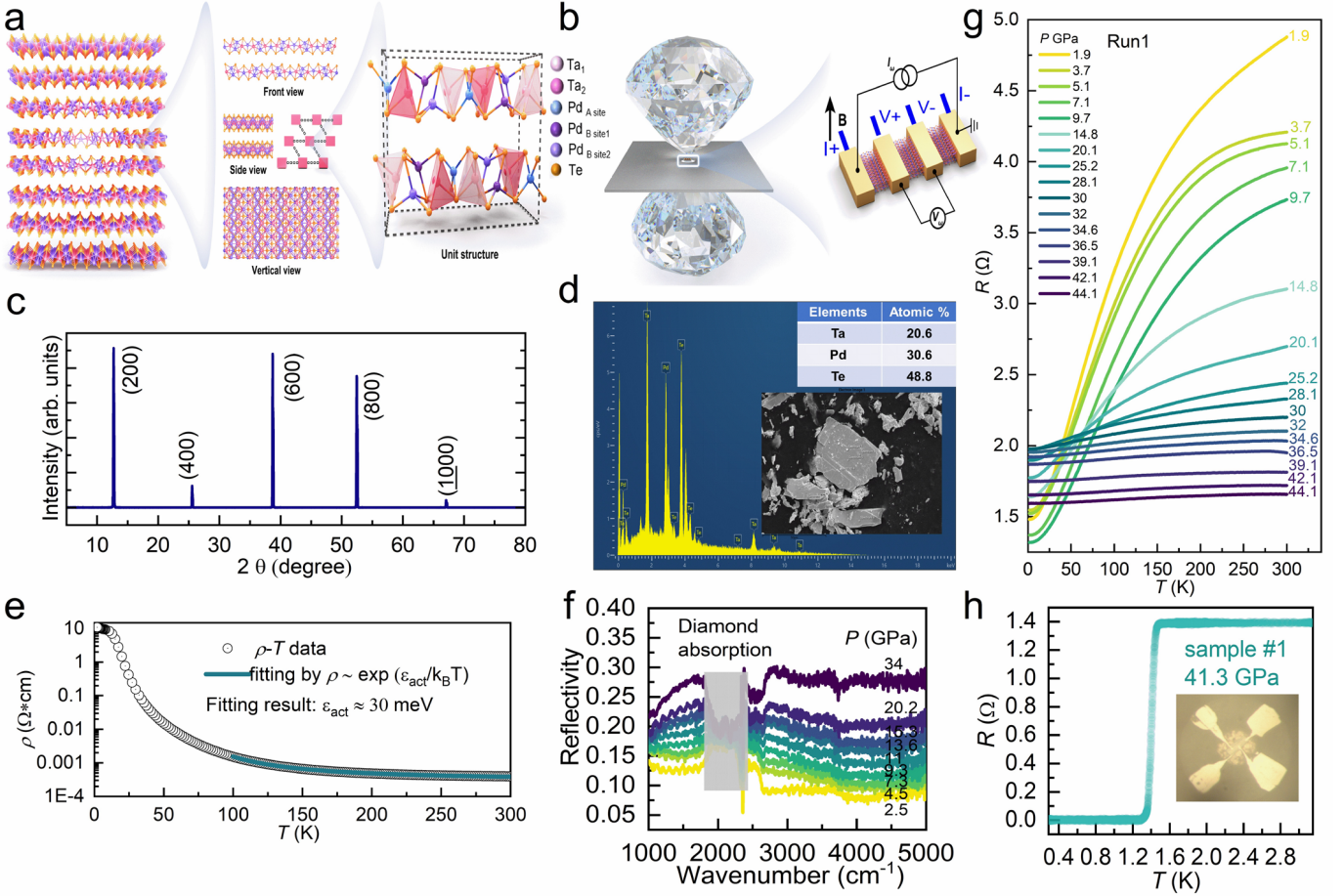

**Figure 1.** Crystal structure characterization and QSH insulator-superconductor transition of $Ta_2Pd_3Te_5$ van der Waals crystal under pressure. (a) Schematic crystal structure with space group *Pnma*. Magenta, violet, and orange balls represent Ta, Pd, and Te atoms, respectively. (b) Schematics for the high-pressure DAC setup using the AC Lock-in amplifier 5640 technique (black wires) and the DC four-probe method (blue wires), the Pt /Au electrodes on diamond culet. (c) XRD spectrum on the flat surface of single crystal. (d) SEM image and elemental contents by EDS. (e) Variation of in-plane resistivity with temperature for single crystals at ambient pressure. The cyan line represents the fitting obtained from the Arrhenius formula. (f) Infrared reflectivity spectra under high pressure (g) Temperature-dependent resistance of $Ta_2Pd_3Te_5$ at elevated pressures up to 44.1 GPa. (h) A novel superconducting phase emerges under 41.3 GPa, and the inset shows the respective optical microscope image.

In order to investigate the electrical transport characteristics of bulk $Ta_2Pd_3Te_5$, we conducted four-probe DC measurements (blue wires) of its temperature-dependent resistivity $R(T)$ up to 45 GPa along the *a*-axis, using diamond anvil cell (DAC) apparatus, as illustrated in Figure 1b. Resistance measurements in two runs reveal that the QSH state preserves within the ultra-narrow pressure range (Figure S3a, c), accompanied by metallization arising from the inherent narrow-gap feature (Figure 1g). Additionally, the $Ta_2Pd_3Te_5$ bandgap could be well-tuned by means of Ti and W chemical doping.[43] When the pressure exceeds 2 GPa, the $R(T)$ curves exhibit semiconductor-to-metal transition, which is similar to the $R(T)$ behavior observed in $Ta_2Ni_3Te_5$.[41] Pressure-dependent enhanced HP-infrared reflectivity indicates the associated increase in carrier concentration under pressure (Figure 1f). Furthermore, it should be noted that the optical oscillation amplitude in the IR reflectance signal diminished when subjected to a pressure of 34 GPa.[44]

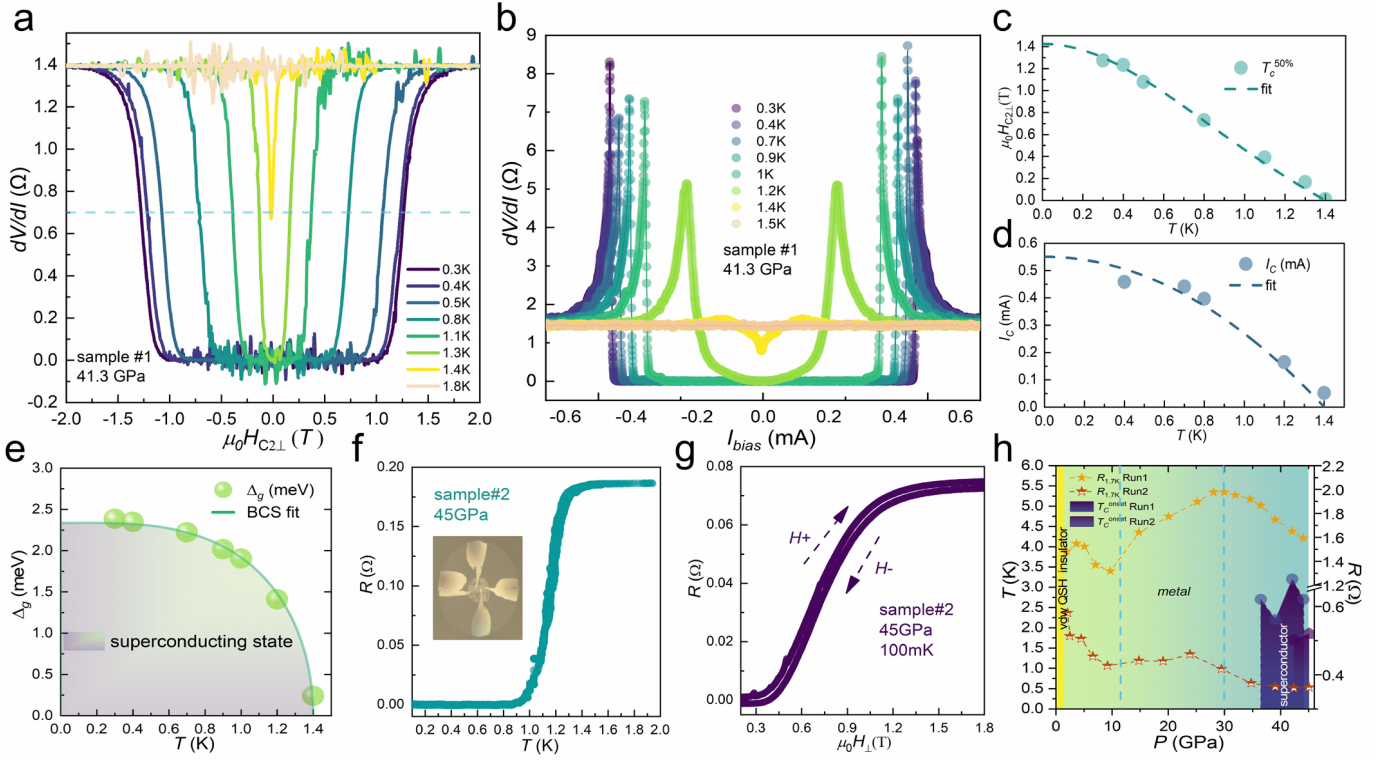

**Figure 2.** Emergent superconductivity in compressed $Ta_2Pd_3Te_5$ samples. (a) Superconductivity transition of the 2.6-$\mu$m-thick sample #1 under a perpendicular magnetic field ranging from 0.3 K to 1.8 K. (b) Differential resistance at multiple predetermined temperatures. The measurements were conducted using the standard lock-in amplifier technique. (c) Temperature dependence of the upper critical field $\mu_0 H_{C2\perp}$, where $T_c$ is defined as 50% drop of the normal state resistance. The dashed line is fitting to the Ginzburg-Landau theory. (d) Temperature dependence of the critical current $I_c$ extracted from d$V$/d$I$ curves. (e) The relationship between temperature and superconducting gap. The olive balls are obtained from the d$V$/d$I$ curve, and the fitting result is based on the Bardeen-Cooper-Schrieffer (BCS) model. The navy-blue shaded area corresponds to the superconducting state. (f) High pressure $R(T)$ shows superconductivity in the sample #2 at 45 GPa. (g) Magnetic field dependence of the resistance with the 5.3-$\mu$m-thick $Ta_2Pd_3Te_5$ sample #2 at 0.1 K. (h) Temperature-pressure phase diagram for the entire pressure region, and the right panel shows specific resistances as a function of pressure at 1.7K.

When the pressure reaches 36 GPa, a slight upturn in the temperature-dependent resistance arises, and pressure-induced superconducting with $T_C^{onset}$ = 1.5 K persists up to 44.1 GPa. An enlarged view of $R(T)$ is depicted in Figure 1h, exhibiting the superconductivity transition at ~1.5 K terminating with the zero resistance at ~1.3 K. $T_c^{onset}$ is defined as the intersection of the extension of normal state resistance and the falling slope of $R(T)$. Upon further increasing the pressure, the $R_{1.7\,K}$ and $R_{300\,K}$ values tend to saturation and are less sensitive to pressure, indicating that $Ta_2Pd_3Te_5$ evolves into the metallic state from semi metallic state (Figure S3b, d).

To enhance our understanding of the modified electronic states contributing to the emerging superconductivity, we conducted supplementary investigations using high-pressure transport measurements. Subsequently, the differential resistance measurements were examined in a quasi-four-probe configuration using the AC lock-in technique (black wires), as shown in Figure 1b. The experimental procedure and additional information can be found in References 45 and 46.[45,46] These measurements include the field dependence of differential resistance $R(H)$ (**Figure 2**a) and critical current dependent differential resistance (Figure 2b). The evolution of $\mu_0 H_{C2}$ is quantified from the 50% drop noted in the $R(H)$ curves. For $P$ = 41.3 GPa, Figure 2a shows the temperature-dependent resistivity $R(H)$ of $Ta_2Pd_3Te_5$ at various temperatures (0.3K~1.8 K), with the magnetic field applied along the $a$-axis. It is evident from the data that the superconducting upper critical field $\mu_0 H_{C2}$ diminishes gradually with increasing temperature. The relationship between the critical field $H_{c2}$ and $T_c$ can be well-fitted by using the Ginzburg-Landau (GL) formula, as illustrated in Figure 2c:[47]

$$\mu_0 H_{C2}(T) = \mu_0 H_{C2}(0)[1-(T/T_C)^2/1+(T/T_C)^2]. \tag{1}$$

The upper critical field, $\mu_0 H_{C2}(0)$, is estimated to be 1.425 T at 41.3 GPa. The in-plane GL superconducting coherence length is further determined to be $\xi(0) = 152$ Å by using the GL equation $\mu_0 H_{C2}(0) = \Phi_0/2\pi\xi(0)^2$, where $\Phi_0 = h/2e$ represents the magnetic flux quantum. The differential resistance $dV/dI$ of the 62-$\mu$m-length $Ta_2Pd_3Te_5$ sample was measured under the effect of bias current, $I_{Bias}$, at multiple selected temperatures from 0.3 K to 1.5 K (Figure 2b). The well-developed U-shape of the d$V/dI$ curves gradually narrow with increasing temperature and eventually become flat around 1.5 K, which is consistent with DC resistance outcomes. Additionally, the temperature dependency of the critical current $I_C$ was extracted from the d$V/dI$ curves, and the value of $I_C(0) = 0.55$ mA can be well-deduced by using the empirical formula (Figure 2d):[48,49]

$$I_C(T) = I_C(0)(1-(T/T_C)^2). \tag{2}$$

The extracted superconducting gap values $\Delta_g(T)$ are plotted in Figure 2e as a function of the reduced temperature $T/T_c$, which follows the Bardeen-Cooper-Schrieffer (BCS) theory,[48]

$$\Delta_g(T) = \Delta_g(0)\,tanh\,(\pi/2\sqrt{(T_C-T)/T}). \tag{3}$$

The obtained zero-temperature superconducting energy gap is $\Delta_0 \sim 2.335$ meV and $2\Delta_0/k_B T_C = 3.8665$. Moreover, in order to verify the repeatability of the experiment, the zero-resistance behavior is also observed in the second sample from different flakes under 45GPa (Figure 2f). A trend of increasing resistance values upon a magnetic field at 100 mK has been observed, and the superconducting state completely vanishes at 1.2 T (Figure 2g). The essential variables, namely temperature (*T*) and pressure (*P*), are concisely summarized, and graphically represented in the *T-P* phase diagram, as illustrated in Figure 2h. The insights gained from high-pressure differential resistance present an innovative approach to directly investigate the superconducting gap properties and underlying mechanisms behind these superconducting states under pressure within DAC setup.

## 2.2 Structure evolution and lattice dynamical properties of $Ta_2Pd_3Te_5$ under pressure

To elucidate whether the pressure-induced transitions to the superconducting phase were triggered by crystal structure changes, *in-situ* high-pressure synchrotron X-ray diffraction and Raman spectroscopy measurements were performed to determine the structure evolution of $Ta_2Pd_3Te_5$. The XRD data have been collected with X-ray wavelength of $\lambda = 0.6199$Å (**Figure 3**a), and the Bragg peaks can be accurately indexed by the space group *Pnma* at a low pressure of 3.8 GPa (Figure S5a). When the pressure was increased to 45 GPa, the Bragg peaks shifted to higher angles, indicating that the lattice parameters were compressed, while the absence of discernible peaks indicate that no structural phase transition occurred. Although the peaks broaden significantly beyond 23.7 GPa, probably due to powder anisotropy, direct identification of the phases becomes challenging. Nonetheless, we efficiently refine the XRD patterns within the pressure range of 1.7~45 GPa using the orthorhombic structure for 10.7 GPa, 23.7 GPa, and 42.8 GPa, as presented in Figure 3c (more details shown in Figure S5). The broadening of the XRD peaks were reversible after decompression (more details shown in Figure S4), clearly excluding the amorphization of $Ta_2Pd_3Te_5$ crystal under high pressure. The XRD data were refined using the LeBail method with the FullProf software, and the lattice parameters obtained are elaborated in Table S3. The pressure dependence of relative axial compressibility is depicted in Figure 3g. The *a*-axis was more compressible, which experienced the most significant level of 6% due to weak vdW interaction between the layers stacked along the *a*-direction. The volume was modeled as a function of pressure using the third-order Birch-Murnaghan equation of state (EOS)[50]

$$P(V) = (3/2)B_0\left[(V_0/V)^{7/3} - (V_0/V)^{5/3}\right] \cdot \left\{1 + (3/4)(B_0' - 4)\left[(V_0/V)^{2/3} - 1\right]\right\}, \qquad (4)$$

where $B_0$ is the bulk modulus, $B_0'$ is the pressure derivative of $B_0$, and $V_0$ is the volume at ambient pressure, and the results are shown in Figure 3g. The cell volume under ambient conditions was determined to be $V_0 = 932(6)$ Å$^3$, and the $B_0$ was calculated to be 188 GPa with $B_0' = 8$ fixed. Low-dimensional transition metal chalcogenides possessing 2D structure, exemplified by $TaPdSe_6$, $TiTe_2$, and $HfS_2$,[51–53] exhibited larger $B_0'$ values, indicating susceptibility to volume compression under pressure.

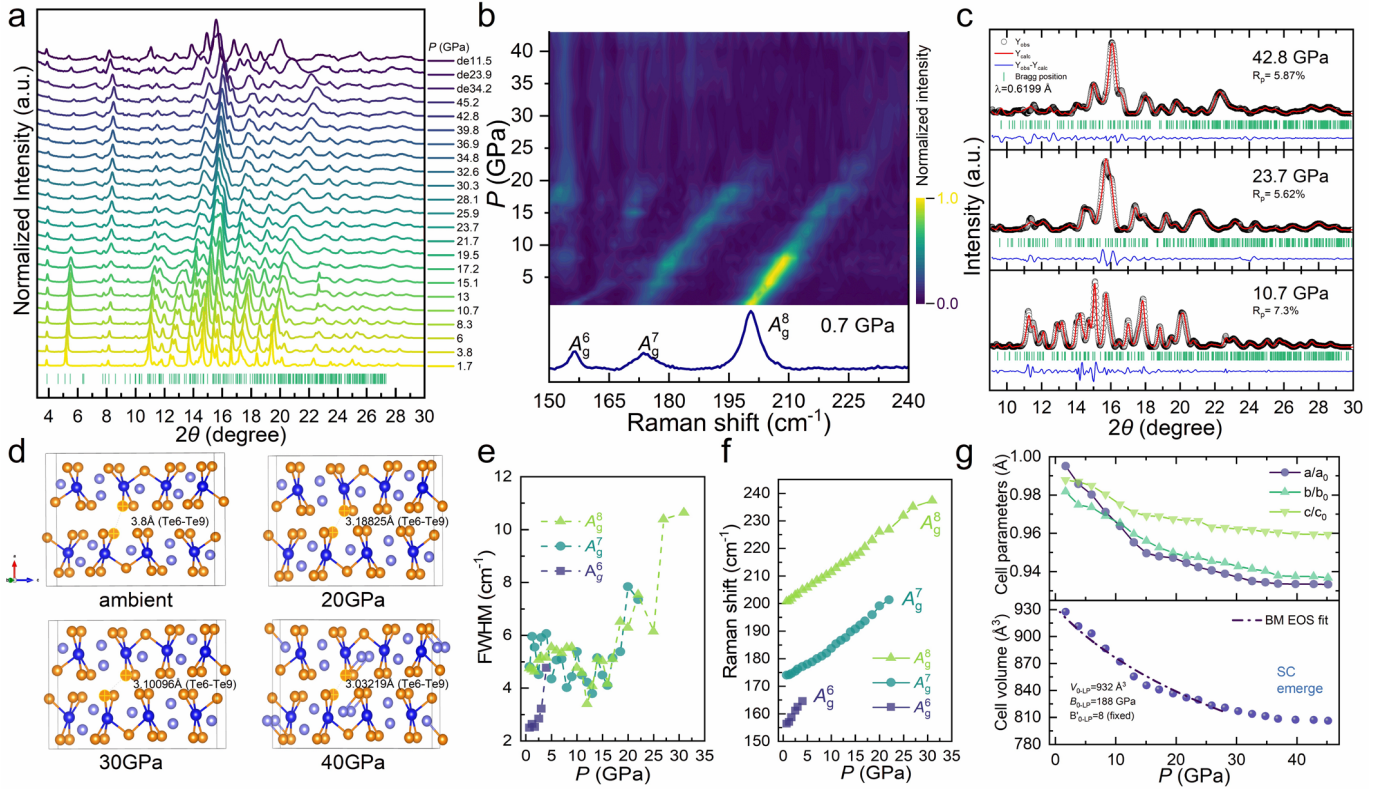

**Figure 3.** Structure evolution and lattice dynamical properties of $Ta_2Pd_3Te_5$ under pressure. (a) *In-situ* XRD patterns plotted as a function of pressure and diffraction angle. (b) Color mapping of normalized Raman spectra plotted as a function of pressure and Raman vibration frequency at room temperature. The bottom panel shows an individual Raman spectrum at 0.7 GPa. (c) Refined XRD patterns at 10.7GPa, 23.7GPa, and 42.8GPa. The synchrotron X-ray wavelength is $\lambda = 0.6199$Å. (d) Density Functional Theory (DFT) predicted the $Ta_2Pd_3Te_5$ crystal structural evolution at various pressures. (e) FWHM of the Raman modes under pressure. (f) Relationship between pressure and the wavevectors of vibration modes is characterized by a monotonic increase with metallization above 20 GPa. (g) Normalized compression ratio of *a*, *b*, and *c* axis. Unit-cell-volume evolution with pressure extracted from XRD spectra. The navy-blue dashed line is a fit to the third-order Birch-Murnaghan equation of state.

We further carried out *in-situ* Raman spectroscopy measurements to explore lattice dynamical properties under pressure. The $Ta_2Pd_3Te_5$ crystal occupies orthorhombic SG *Pnma* (No. 62), where the Wyckoff positions are 4c for all atoms. Group theory analysis predicts the existence of $20A_g + 10B_{1g} + 20B_{2g} + 10B_{3g}$ Raman-active modes in $Ta_2Pd_3Te_5$. The analogous compound $Ta_2Pd_3Se_8$ possesses the highly anisotropic features of the heterostructure, enabling its selective detection by linear polarized light,[54] and $Ta_2Pd_3Te_5$ also presents distinctive anisotropic attributes. The discussion of observed Raman spectra is focused on the following prominent active phonon vibrations at ambient pressure: the $A_g^6$ mode at ~156.72 cm$^{-1}$, the $A_g^7$ mode at ~174.59 cm$^{-1}$, and the most intense peak $A_g^8$ mode around 198.52 cm$^{-1}$ (Figure 3b, bottom panel). The mapping of normalized Raman spectra in Figure 3b shows the evolution of these phonon modes with pressure (more details are shown in Supporting Information Figure S6, S7, and S8). The $A_g^6$ mode ~156.72 cm$^{-1}$ vanish at low pressure~ 2.5 GPa, whereas the modes of the other two peaks, $A_g^7$ and $A_g^8$, initially exhibit clear blue shift, and later transforming into two broader peaks around 201 cm$^{-1}$ and 216 cm$^{-1}$, respectively (Figure 3e and 3f). As the pressure continues to rise beyond 20 GPa, the intensities of these two broader peaks rapidly diminish, reaching the background noise level. The notable transformation in the Raman spectra presents supplementary confirmation of the isostructural phase transition induced by pressure, in accordance with our theoretical structure predictions (Figure 3d) and XRD measurements.

**2.3 Theoretical calculations on electronic band structures and potential crystal structures**

To gain insights into the effects of applied pressure on electronic and structural properties, we conducted first-principles calculations up to 40 GPa, enabling us to analyze how pressure modifies the properties of bulk Ta$_2$Pd$_3$Te$_5$. The calculations presented in **Figure 4** demonstrate significant alterations in both the electronic bands and density of states (DOS) of Ta$_2$Pd$_3$Te$_5$ when subjected to pressure. At ambient pressure, Ta$_2$Pd$_3$Te$_5$ exhibits semi-metallic nature with a narrow energy gap located at the center (T, Z, Γ point) of its Brillouin zone (Figure 4a). The Fermi energy ($E_F$) levels may reside within the DOS valley at pressures below 10 GPa as depicted in Figure S10. Above 10 GPa, the conduction bands of Γ point cross at $E_F$ (Figure 4b), and the most prominent transformation observed is the transition from the semi-metallic to metallic state, which occurs at the critical pressure 18 GPa (Figure S12c and S12d). Under 25 GPa, the valence, and conduction bands are further crossed near the Fermi level, the conduction band located at R, X, and S point, experiencing a downward repositioning (Figure 4c). The states near $E_F$ are primarily contributed by Ta-5$d$ (27.83%) and partially by Pd-4$d$ (13.47%) and Te-5$p$ (16.36%) electrons (Figure S11), exhibiting significantly enhanced DOS at the $E_F$ compared to 10 GPa. The results aligned with the dramatic changes obtained from Raman data, suggesting the potential presence of bulk superconductivity in Ta$_2$Pd$_3$Te$_5$ (Figure S10). As the pressure further increasing, the consistent trend of bands shifting between the conduction and valence bands persists and the band gap has completely vanished (Figure S12e and S12f). To enhance our understanding of the high-pressure phases of Ta$_2$Pd$_3$Te$_5$, we employed crystal structure prediction techniques to explore potential structures for elevated pressures up to 50 GPa. The plotted enthalpy-pressure curves in Figure 4d highlight the most promising candidates identified by our structural prediction. The calculated energetic data establish the following stability sequence: *Pnma* phase < *P2$_1$/m* phase < *Pnnm* phase at equilibrium lattice parameters. Upon compression, the *Pnma* structure maintain the lowest entropy phase as the most thermodynamically favorable, corroborating our experimental XRD observations. The crystal structures of *P2$_1$/m* (monoclinic phase) and *Pnnm* (interlayer slip phase) at 30GPa are plotted in Figure 4d. The underlying dynamic process for the phases transformation from ambient phase toward denser phase for all the relevant candidates are illustrated in Figure S13, S16, and S19. The lattice constants identified from the theoretical prediction approach can be found in Figure S14, S17 and S20. To validate the dynamic stability of monoclinic structure, we performed calculations on phonon band structures and partial density of states (PDOS) within $P2_1/m$ symmetry at 40 GPa. Figure 4e illustrates the results, indicating the absence of any imaginary frequencies across the entire Brillouin zone. The high-frequency modes centered around 240 cm$^{-1}$ primarily originate from Pd and Te atoms, while the low-frequency ones are predominantly contributed by Te atoms. To establish the experimental relevance of the high-pressure phase discovered by theoretical calculations, we compared its simulated XRD spectra with those of the $P2_1/m$ phase and $Pnnm$ phase at 20 GPa. The simulated XRD patterns provide an excellent match to the previously unexplained diffraction peaks identified around 15.4° and 15.8° in our experiments, as shown in Figure 4f. Figure S15, S18 and S21 depict the additional simulated XRD patterns for the $Pnma$ phase, $P2_1/m$ phase and $Pnnm$ phase under high pressure.

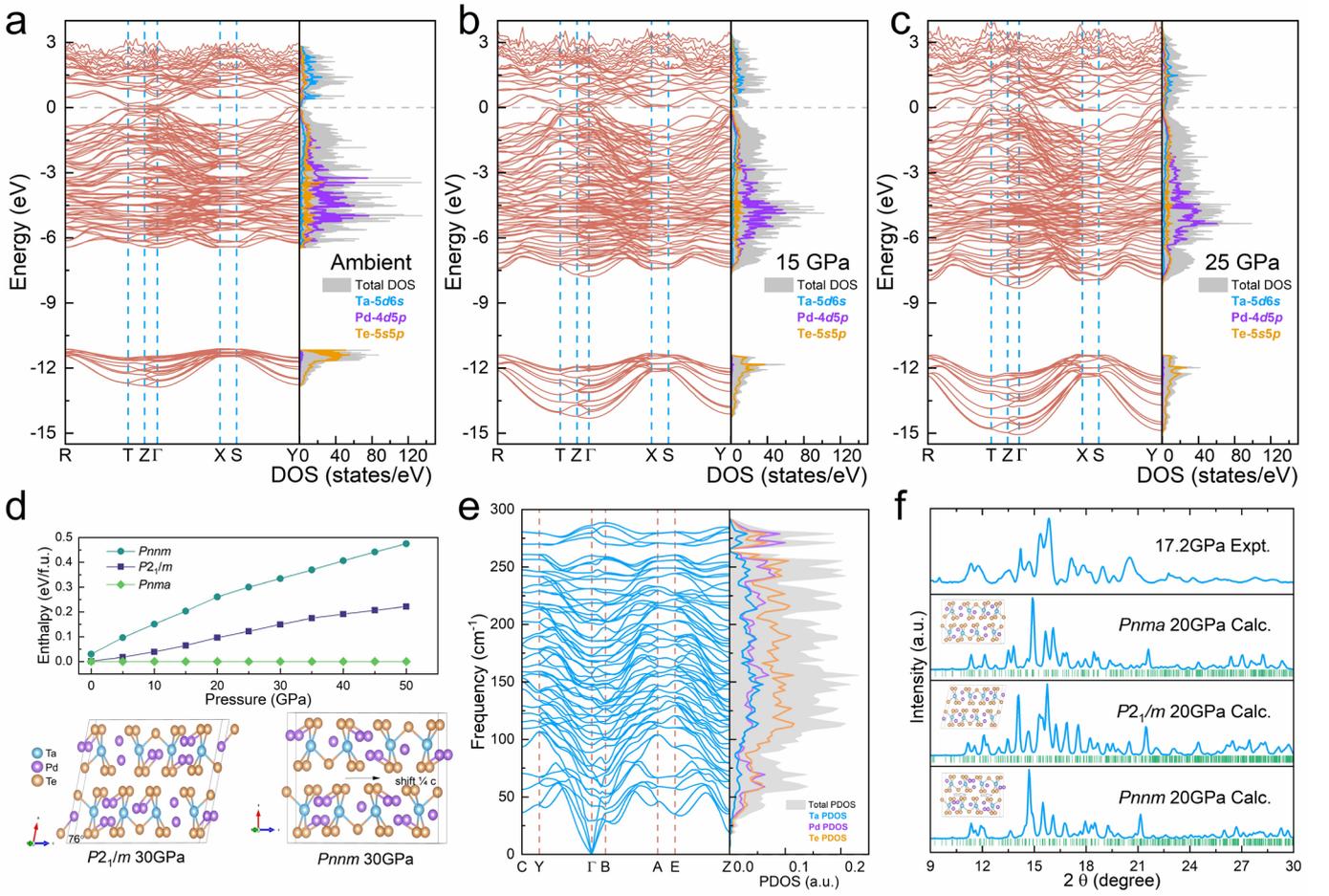

**Figure 4.** Calculated electronic band structures and density of states (DOS) at (a) ambient pressure, (b) 15 GPa, and (c) 25 GPa. (d) Enthalpy calculation of possible stable phases relative to the $Pnma$ phase as a function of pressure. Crystal structure of the high-pressure phases of Ta$_2$Pd$_3$Te$_5$, $P2_1/m$ and $Pnnm$. (e) Phonon band structures and phonon density of states (PDOS) for the monoclinic $P2_1/m$ phase at 40 GPa. (f) Experimental synchrotron XRD patterns for Ta$_2$Pd$_3$Te$_5$ and simulated XRD patterns for the $Pnma$ phase, $P2_1/m$ phase, and $Pnnm$ phase at 20 GPa (shown for comparison).

## 3. Conclusion

In this work, we employed a combination of electrical transport measurements, theoretical calculations, optical infrared reflectance, *in-situ* high-pressure synchrotron X-ray diffraction (HP-XRD), and Raman spectroscopy to offer compelling proof of the substantial impact of strain engineering on topological bands. Our electrical resistance measurements confirmed the metallization and emergence of superconductivity. The metallization was further supported by the HP-Raman spectroscopy and HP-IR reflexivity measurements. When pressure exceeded 20 GPa, the Raman signal diminished significantly, and there was a pronounced rise in IR reflectivity, signifying a rapid increase in carriers. Moreover, the patterns derived from synchrotron radiation X-ray diffraction, which did not identify any new peaks, clearly illustrates isostructural modifications at low-pressures and 20 GPa, modifications that correspond to changes in the spatial arrangement of atoms. Furthermore, first-principles calculations revealed remarkable changes in electronic properties, as well as the trend of band shifting between the conduction and valence bands under pressure. We also reconstructed and discussed the potentials of the compressed interlayer slip phase (*Pnnm*) and monoclinic phase (*P*2$_1$/*m*) *vs* the pristine phase (*Pnma*). Interestingly, the pressure-dependent free energy $E(P)$ simulation showed that the energy across the entire span of pressure levels for the pristine robust structure (*Pnma*) remained at tis minimum value, as evidenced by the consistency between computed XRD outcomes and actual observations. As pressure increased, the interlayer atomic positions gradually approached each other, inducing a phase transition from the initially loose insulating vdW phase to a denser high-coordination phase, eventually stabilizing the interlayer above the pressure of 40 GPa. The quantum topological phase transition attributes stemming from the strain-induced accelerated the enrichment mechanism of DOS, which involves a reversible volume compression of the topological insulator Ta$_2$Pd$_3$Te$_5$. Our findings established a feasible framework for the practical manipulation of both superconductivity and band topology in 2D vdW materials, which holds significant promise for applications such as superconducting diodes and topological quantum computing.

## 4. Experimental Section

*Sample Synthesis and Characterizations.* Ta$_2$Pd$_3$Te$_5$ single crystals were synthesized using the self-flux method. A molar ratio of Ta:Pd:Te = 2:4.5:7.5 mixture was placed in alumina crucible and sealed in evacuated quartz tube. The tube was gradually heated to 950°C for over 10 h and held at this temperature for two days. Subsequently, the tube was slowly cooled to 800°C at a rate of 0.5°C/h. Finally, the extra flux was removed by centrifugation at 800°C. Crystalline structure analysis was conducted via single-crystal XRD using Mo Kα radiation ($\lambda$ = 0.71073 Å) at 273 K. Electrical resistivity measurements were performed under ambient pressure using a physical property measurement system (Quantum Design, Inc.)

*In-Situ High-Pressure XRD/Raman Spectroscopy/IR Reflectance Spectroscopy.*

In situ high-pressure synchrotron powder XRD measurements were performed at the 4W2 station of the Beijing Synchrotron Radiation facility (BSRF) with the wavelength 0.6199 Å. A symmetric diamond anvil cell (DAC) with anvil culet diameters of 300 $\mu$m was employed. The T301 stainless steel gasket was perforated to generate hole of ~80 $\mu$m providing the sample high pressure chamber. The pressure was determined by the ruby fluorescence technique, with silicone oil serving as the pressure transmitting medium. The sample-to-detector distance and other detector parameters were calibrated using the $CeO_2$ standard. The raw 2D XRD images were integrated into one-dimensional intensity *vs.* 2$\theta$ angles using the DIOPTAS program,[55] and the patterns were refined by the Le Bail method[56] using the FullProf software. High-pressure Raman spectroscopy was conducted using Horiba Scientific LabRAM HR Evolution confocal Raman micro-spectrometer in the backscattering configuration. The excitation was generated by a diode-pumped solid-state laser operating at 532 nm, with a focused beam diameter of 2.5 $\mu$m, and the Raman signals were separated using 600 g/mm$^{-1}$ grating. KBr was employed as the medium for transmitting pressure. HP-IR reflectance spectra were measured using a Bruker VERTEX 70v instrument with the wavenumber range of 500−6000 cm$^{-1}$.

*High-Pressure Electrical Resistance Measurements.*

The HP-electrical resistance of the sample loaded in Be-Cu DAC was measured using the four-probe method at selected pressures up to 45.1 GPa and temperatures ranging from 1.6 K to 300 K. A rhenium gasket was preindented to ~30 $\mu$m in thickness, and a hole of 150 $\mu$m in diameter was drilled through the gasket center. To shield the electrodes (platinum) from the gasket, the insulating cubic boron nitride (c-BN) was placed as the filler for pressurization. The sample chamber was then created by drilling a hole in the middle of the c-BN layer with a diameter of 90 $\mu$m, and KBr was used as the pressure medium. A multifunctional measurement equipment, which included a magnet Dewar flask (9 T C-MAG, Cryomagnetic, Inc.), a signal source (Tektronix 6221-Keithley), and a nanovoltmeter (Keithley-2182), was used to test the low-temperature electrical resistance of $Ta_2Pd_3Te_5$ under high pressure. An Oxford cryostat with a $^3$He insert was used to cool the whole DAC setup down to 0.3 K, and a magnetic field was applied along the crystal *a*-axis up to 1000 Oe to suppress superconductivity.

*Electronic Structure Computational Methods:*

Our density functional theory (DFT) calculations were performed using the Vienna *ab initio* simulation package (VASP)[57] with the all-electron projector augmented wave method.[58] The Perdew-Burke-Ernzerhof generalized gradient approximation revised for solids (PBEsol-GGA)[59] exchange-correlation functional was adopted to evaluate the structural stability of $Ta_2Pd_3Te_5$ under pressure. The valence states $5d^46s^1$ for Ta, $4d^95p^1$ for Pd, and $5s^25p^4$ for Te were used with the energy cutoff of 500 eV for the plane wave basis set. To identify the possible new structures, we have performed a structure search based on a layer-by-layer slip reconstruction mechanism. This mechanism has been employed in the structure search in graphite[60] based on a modified climbing image nudged elastic band method (CI-NEB)[61] for solid-state materials[62] with the cell and atomic position optimized under a wide pressure range of 5–40 GPa. Phonon calculations were performed using the MedeA-Phonon code[63] for the new $Ta_2Pd_3Te_5$ structures under pressure. The geometries were optimized until the remaining atomic forces were less than $10^{-2}$ eV/Å, and the energy convergence criterion was set to $10^{-6}$ eV. Electronic band structures and density of states were calculated using the modified Becke-Johnson (mBJ)[64] functional.

**Supporting Information**

Supporting Information is available from the Wiley Online Library or from the author.

**Acknowledgments**

This work was supported by the National Key R&D Program of China (Grants No. 2016YFA0401503, No. 2017YFA0302900, No. 2018YFA0305700, No. 2020YFA0711502, and No. 2021YFA1400300, the National Natural Science Foundation of China (Grants No. 11921004, No. 92263202, No. 12374020, and No.11820101003), the Strategic Priority Research Program and Key Research Program of Frontier Sciences of the Chinese Academy of Sciences (Grants No. XDB33000000, No. XDB25000000, and No. QYZDBSSW-SLH013), the Youth Innovation Promotion Association of Chinese Academy of Sciences (No. Y202003), the Beijing Nature Science Foundation (No. 2202059), the Guangdong Innovative & Entrepreneurial Research Team Program (No. 2016ZT06C279), and the Shenzhen Peacock Plan (No. KQTD2016053019134356). The in situ XRD measurements were performed at 4W2 High Pressure Station, Beijing Synchrotron Radiation Facility (BSRF), which is supported by the Chinese Academy of Sciences (Grants No. KJCX2-SW-N20 and No. KJCX2-SW-N03). This work was partially carried out at the high-pressure synergetic measurement station of Synergic Extreme Condition User Facility.

**Conflict of Interest**

The authors declare no conflict of interest.

**Data Availability Statement**

The data that support the findings of this study are available from the corresponding author upon reasonable request.